# Drastic suppression of the superconductivity of LaFeAsO$_{0.85}$ by a nonmagnetic impurity

Y. F. Guo,[1,2,*] Y. G. Shi,[1,2] S. Yu,[3] A. A. Belik,[1,2] Y. Matsushita,[4] M. Tanaka,[4] Y. Katsuya,[5] K. Kobayashi,[4] I. Nowik,[6] I. Felner,[6] V. P. S. Awana,[1,7] K. Yamaura,[2,3] E. Takayama-Muromachi,[1,2,3]

[1] International Center for Materials Nanoarchitectonics (MANA), National Institute for Materials Science, Tsukuba, Ibaraki 305-0044, Japan

[2] JST, Transformative Research-Project on Iron Pnictides (TRIP), 5 Sanbancho, chiyoda-ku, Tokyo 102-0075, Japan

[3] Superconducting Materials Center, National Institute for Materials Science, 1-1 Namiki, Tsukuba, 305-0044 Ibaraki, Japan

[4] NIMS Beamline Station at SPring-8, National Institute for Materials Science, 1-1-1 Kouto, Sayo-cho, Sayo-gun, Hyogo 679-5148, Japan

[5] SPring-8 Service Co. Ltd., 1-1-1 Kouto, Sayo-cho, Sayo-gun, Hyogo 679-5148, Japan

[6] Racah Institute of Physics, The Hebrew University, Jerusalem, 91904, Israel

[7] National Physical Laboratory, Dr. K.S. Krishnan Marg, New Delhi 110 012, India

We observed a drastic $T_c$ suppression caused by no more than 3 at.% of Zn doped to the optimized superconductor LaFeAsO$_{0.85}$ ($T_c$ = 26 K). The electrical resistivity and magnetic susceptibility measurements suggested that it is likely due to impurity scatterings rather than losing the metallic nature, seeming to support the $s_{+-}$ pairing model proposed for the Fe pnictide superconductor.

Discovery of the superconductivity (SC) in the quaternary oxyarsenide LaFeAsO$_{1-\delta}$F$_y$ stimulated tremendous activities in condensed-matter physics [1]. Consequently, the superconducting transition temperature ($T_c$) went over 50 K by replacing La to other rare-earth element, and moreover, a variety of Fe pnictides containing the Fe$_2$As$_2$ layer in each structure have been proved to become superconducting, e.g., $A$Fe$_2$As$_2$ ($A$ = alkaline earth) [2], $A$FeAsF [3], and Sr$_4$Sc$_2$Fe$_2$P$_2$O$_6$ [4]. The Fe-based superconductors also attract attentions from the application side because of their remarkably high upper-critical field [5].

The SC in the Fe pnictides emerges after the spin-density wave (SDW) is suppressed by a chemical or a physical carrier doping, being analogical with features of the high-$T_c$ Cu oxides [6]. The Fe$_2$As$_2$ layer is believed to play a decisive role in establishing SC as does the CuO$_2$ layer [7]. The facts imply common physics in part between the Fe pnictides and the Cu oxides. However, it is widely believed that the superconducting paring symmetry is different in nature between the 2 superconducting systems. A fully-gapped model ($s$-wave) was suggested for the Fe pnictide [8,9], while a $d$-wave model for the Cu oxide. It is thus important to investigate nature of the pairing symmetry of the Fe pnictide in order to further promote establishing the mechanism of the SC.

Based on the Fermi surface topology, a fully-gapped sign-reversing $s$-wave (so-called $s_{+-}$ wave) model was suggested for the Fe pnictide by at least 2 independent groups, and it indeed looks compelling [10-12]. However, a recent study cast a doubt for the model based on the theoretical prediction that the $s_{+-}$ wave SC should be quite fragile against an impurity [13,14]. In the experimental side, the SC of the Fe pnictide was found robust against a variety of doped impurities such as Co, Ni, Ru, Rh, Pd, and Ir [15-18], in striking contrast with the prediction. In Refs. 13 and 14, it is however stated that the $s_{+-}$ wave model can be possibly consistent with the robust SC if certain conditions such as a small impurity potential (<<1 eV) and a large potential radius are effective in the SC. In turn, only the way in impurity studies to support the $s_{+-}$ wave model is to confirm the fragile SC against impurity scatterings. To the best of our knowledge, such the fragile

SC is not observed yet in the Fe pnictide.

As far as we know, only a one experimental result was reported regarding the fragile SC: it is for $Ln$Fe$_{1-y}$V$_y$AsO$_{1-\delta}$F$_x$ ($Ln$ = La and Nd) [19]. However, it is stated that the observed $T_c$ suppression is not due to impurities scatterings but rather due to losing conductivity. In this paper, we report a drastic suppression of $T_c$ caused by a minimal level of Zn substitution (below 3 atomic %) to the optimized superconductor LaFeAsO$_{0.85}$ [20]. On the contrary to the V study, the magnetic and electrical data clearly indicate that it is not due to losing the conductivity. We will discuss the role of Zn in the $T_c$ suppression, and show that the observation well accords with the prediction from the $s_{+-}$ wave model.

Polycrystalline samples of LaFe$_{1-x}$Zn$_x$AsO$_{0.85}$ ($x$ = 0, 0.005, 0.01, 0.015, 0.02, 0.03) and LaFeAsO$_{1-\delta}$ ($\delta$ = 0, 0.12, 0.15, 0.22) were prepared by a solid-state method from LaAs (lab-made), Fe$_2$O$_3$ (3N, Furuuchi Chem. Co.), ZnO (3N, Wako), and Fe (3N, 100 mesh, Rare Metallic Co.). The LaAs precursor was prepared from La pieces (3N, Nilaco Co.) and As powder (5N, High Purity Chem.). 1 at.% excess As was added. The La-As mixture was heated in an evacuated quartz tube at 500 ºC for 20 h, followed by quenching to the room temperature, grinding, and re-heating at 850 ºC for 10 h in an evacuated quartz tube.

The starting mixture each was loaded into a gold capsule with an h-BN inner, which was pre-heated at ~2000 ºC for 1 h in nitrogen to remove possible boric oxides. The capsule was heated at 1300 ºC for 2 h in a belt-type press, which is capable of maintaining 6 GPa during the heating, followed by quenching to room temperature before releasing the pressure. To gain the chemical uniformity of the sample, the pellet was carefully ground and re-heated in the same way. The $\delta$ = 0 sample was synthesized in an evacuated quartz tube at 1100 ºC for 30 h without using the press.

The phase purity was checked by a powder x-ray diffraction (XRD) method in a diffractometer, RINT2200V/PC, Rigaku, using Cu-K$\alpha$ radiation. Selected samples of LaFeAsO$_{0.85}$ and LaFe$_{0.99}$Zn$_{0.01}$AsO$_{0.85}$ were further investigated by a synchrotron X-ray diffraction (SXRD) method and a Mössbauer spectroscopy. The SXRD measurement was conducted at $\lambda$ = 0.652973 Å in a large Debye-Scherrer camera at the BL15XU beam line of SPring-8. The sample capillary, Lindenmann glass, was rotated during the measurement. The Rietveld analysis was carried out by using RIETAN-2000 [21]. The Mössbauer spectroscopy was carried out at room temperature by using a conventional constant acceleration drive and a 50 mCi $^{57}$Co:Rh source. The experimental spectra were analyzed by a least-squares fit procedure. The velocity calibration and isomer shift (IS) zero are those of $\alpha$-Fe measured at room temperature.

The magnetic susceptibility ($\chi$) of the samples was measured in a magnetic property measurement system, Quantum Design. Loose powder was cooled to 2 K without applying a magnetic field (zero-field cooling; ZFC), followed by warming in a magnetic field of 10 Oe to 300 K. The sample was again cooled to 2 K in the field (field cooling; FC). The electrical resistivity ($\rho$) was measured in a physical properties measurement system, Quantum Design, by a four-probe method with a constant gauge current of 0.2 mA.

The powder XRD patterns of LaFe$_{1-x}$Z$_x$AsO$_{0.85}$ are shown in Fig. 1a. All peaks were satisfactorily indexed by assuming the ZrCuSiAs-type structure with *P*4/*nmm* as was achieved in Ref. 22, indicating the high quality of the sample. Figs. 1b and 1c show the SXRD-Rietveld analysis results of the selected samples with and without the doped Zn ($x$ = 0 and 0.01) [23]. The relatively small *R* and the smooth difference curves at the bottom of Figs. 1b and 1c each secure the reliability of the refinements. We carefully investigated the refined local structure, since the La-As distance and the Fe-As-Fe angle in the Fe$_2$As$_2$ layer play a crucial role in controlling the effective bandwidth, thereby affecting the SC [24]. The La-As distance is 3.3577(6) at $x$ = 0 and 3.3570(6) at $x$ = 0.01, and the angle is 113.09(9)° and 113.20(9)°, respectively, indicating that the local structure changes quite little over the Zn substitution.

The unit-cell evolution over the Zn substitution was studied by a quantitative analysis of the XRD patterns, as shown in Figs. 2a-2c. Regarding the oxygen-vacant LaFeAsO$_{1-\delta}$, the parameters $a$ and $c$ decrease with increasing $\delta$, reflecting the enhanced Coulomb attractive force between the charged [LaO$_{1-\delta}$]$^{1+2\delta}$ and [FeAs]$^{1-2\delta}$ layers and increased amount of the oxygen vacancies. In contrast to the isotropic change, the Zn substitution results in an anisotropic change: $c$ of LaFe$_{1-x}$Zn$_x$AsO$_{0.85}$ increases (+0.02 % at Zn$_{0.01}$), while $a$ decreases much efficiently (-0.08 %). The different characters indicate that the doped Zn is indeed incorporated in the lattice. Moreover, the change by Zn well follows the Vegard's law up to $x$ of 0.02, suggesting that Zn and Fe are isovalent and the solubility limit of Zn is between $x$ of 0.02 and 0.03. In addition, magnetic $T_c$ (data are shown later) vs. $c/a$ (Fig. 2d) indicates a tight relation between those. It is clear that the Zn substitution is successful in the range $x < 0.03$ under the synthesis condition.

Figs. 3a-3f show the Zn concentration dependence of $\chi$ vs. $T$ for LaFe$_{1-x}$Fe$_x$AsO$_{0.85}$. In Fig. 3a, the Zn-free sample clearly undergoes the superconducting transition at ~26 K as reported elsewhere [20]. Employing the calculated density of 7.85 g/cm$^3$, the magnetic shielding fraction is estimated to be 1.13 (1.00 is expected for the perfect shielding), confirming the high homogeneity of the sample. While, the Meissner fraction (5 K, FC curve) is rather small, being less than 0.1. The magnetic features well coincide with what was observed for the oxygen-deficient systems including TbFeAsO$_{0.85}$ [22]: the peculiarity is likely due to a highly efficient magnetic-flux pin.

With increasing the Zn concentration, $T_c$ drastically goes down in large contrast to what was observed for the Co doped LeFeAsO$_{1-\delta}$ [25]. Unlike Co, no more than 3 at.% of Zn almost completely suppresses the SC. The $T_c$ variation is summarized in the inset to Fig. 3f, showing a drastic $T_c$ suppression. The $T_c$ suppression is highly remarkable because many $d$ elements such as Ni, Ru, Rh, Pd, and Ir doped into the Fe$_2$As$_2$ layer suppress $T_c$ only weakly or even contribute to induce SC [15-18]. In order to secure the remarkable result, we carefully tested a possibility of a run-to-run error in the synthesis by means of repeating the synthesis of all the samples. Two sets of independent data are shown simultaneously in Figs. 3a-3f, confirming clearly that the result is well reproducible. Actually, we repeated all the synthesis 4 times in total, resulted in confirming the reliability.

Figs. 4a and 4b show the temperature dependence of $\rho$ of LaFe$_{1-x}$Zn$_x$AsO$_{0.85}$ and LaFeAsO$_{1-\delta}$, respectively. Regarding the stoichiometric LaFeAsO, $\rho$ at 300 K is ~6 mΩcm, being comparable with the normalized mean free path $k_F l \approx 0.6$ [26]. The $\rho$ of LaFeAsO gently varies with a broad minimum at approximately 220 K and pronouncedly goes down at 150 K, corresponding to the SDW instability [26]. By introducing the oxygen vacancies, the normal-state $\rho$ becomes much smaller, reflecting the increase of the carrier density in the Fe$_2$As$_2$ layer. The residual resistivity ratio (RRR) $\rho(300\ K)/\rho(T_c)$ is ~6 at $\delta = 0.15$ (LaFeAsO$_{0.85}$).

$T_c$ determined from the $\rho$ data steeply shift toward zero by the Zn substitution (Fig. 4a), being entirely consistent with the $T_c$ change in the magnetic data. Regarding the normal stare $\rho$, the quadratic like temperature dependence and the prominent upturn on cooling were observed in the Zn-doped samples (see for example at $x = 0.01$) and RRR is deteriorated somewhat, reflecting enhanced scattering factors. On the other hand, Sato *et al*. found a noticeable increase of the normal state $\rho$ in their V-doped samples, $Ln$Fe$_{1-y}$V$_y$AsO$_{1-x}$F$_x$, which let them propose that the drastic suppression of $T_c$ in the V-doped samples is due to the carrier localization rather than the pair breaking [19]. Thus, we checked the present data on the point. We found that the normal state $\rho$ of LaFe$_{1-x}$Zn$_x$AsO$_{0.85}$ ($x = 0.02$) is much smaller than that of LaFeAsO$_{1-\delta}$ ($\delta = 0.12$, $T_c = 21$ K) in spite of the almost complete disappearance of the bulk SC in the former sample. The electron localization picture is therefore unlikely for the present $T_c$ suppression.

In order to further examine the scattering picture, we carefully measured the oxygen content of the Zn-doped samples by a gravimetric method [27]. The net oxygen content of LaFe$_{1-x}$Zn$_x$AsO$_{0.85}$ was determined to be 0.83, 0.85, 0.84 for $x = 0.005, 0.010$, and $0.020$, respectively, in good agreement with the nominal. It appears that the oxygen content variation is too small to be responsible for the drastic $T_c$ suppression [20,22]. In addition, we studied the Mössbauer effect of the $x = 0$ ($T_c = 26$ K) and 0.01 ($T_c = 10$ K) samples. As shown in Figs. 1b and 1c, both the spectra are identical, suggesting the two samples are of pure and high-quality and the Fe valence is essentially unaltered. In more details, the hyperfine parameters and the IS values are the same and the quadrupole splitting (QSP) is negligibly small [28]. The oxygen content variation is again confirmed fairly small.

Let us focus on the role of the doped Zn. The divalent Zn has the fully occupied configuration $3d^{10}$, in contrast to $Co^{2+}$: $3d^7$ and $Ni^{2+}$: $3d^8$, resulting in a highly localized nature. Indeed, the Zn $3d$ states are located at -8 to -6.5 eV far below the Fermi level in LaZnAsO [29]. It is therefore reasonable that doped Zn in the Fe site does not add itinerant electrons into the $Fe_2As_2$ layer. Since the $T_c$ suppression was efficiently achieved by no more than 3 at.% of Zn at the Fe site without losing the metallic nature and changing the electrons count, it is likely that the doped Zn mostly works as a scattering center in accordance with results of a recent density functional study [30].

In the theoretical studies on a 5 $d$-orbital model, it was found that the Anderson's theorem is violated for the $s_{+-}$ wave state in the Fe pnictide due to strong inter-band impurity scattering, predicting a significant nonmagnetic impurity effect on the SC [13,14]. Thus, the $s_{+-}$ wave model was called in question because of the robust SC observed. Besides, possible certain conditions were suggested to accept the $s_{+-}$ wave model, such as a small impurity potential (<<1 eV) and a large potential radius comparable to the lattice spacing, both of which suppresses momentum scatterings [13,14]. Zn, however, apparently does not meet the conditions.

It is indispensable to mention the Zn study of $LaFeAsO_{1-\delta}F_{0.1}$ [31], which concluded that SC is robust against Zn as well as other impurities. It is stated in Ref. 31 that SC remains almost unperturbed or even enhanced by the Zn substitution (<10 at.%), being strictly contrasting with the present result. Because the improvement of the electrical conductivity by the Zn doping is quite unusual [30], possible additional factors which ameliorate the electronic state should be investigated before we discuss further.

In summary, a drastic $T_c$ suppression from the optimum $T_c$ of 26 K to the lowest limit was observed by a minimal amount of Zn (< 3 at.%) doped to LaFeAsO$_{0.85}$. Although we need more quantitative studies on the impurity potentials and potential radii for various impurities including Zn to reach a solid picture, the $T_c$ suppression is most likely due to strong impurity scatterings rather than due to losing the metallic nature. Hence, the result seems to experimentally support the $s_{+-}$ wave model. The density functional study on the Zn doped LaFeAsO also supports the model as well [30]. Over the impurities studies of the Fe pnictide achieved to date, this is the unique result showing a drastic $T_c$ suppression likely due to impurity scatterings.

We thank Dr. D. J. Singh for valuable discussion. This research was supported in part by World Premier International Research Center (WPI) Initiative on Materials Nanoarchitectonics from MEXT, Japan, Grants-in-Aid for Scientific Research (20360012) from JSPS, Japan.

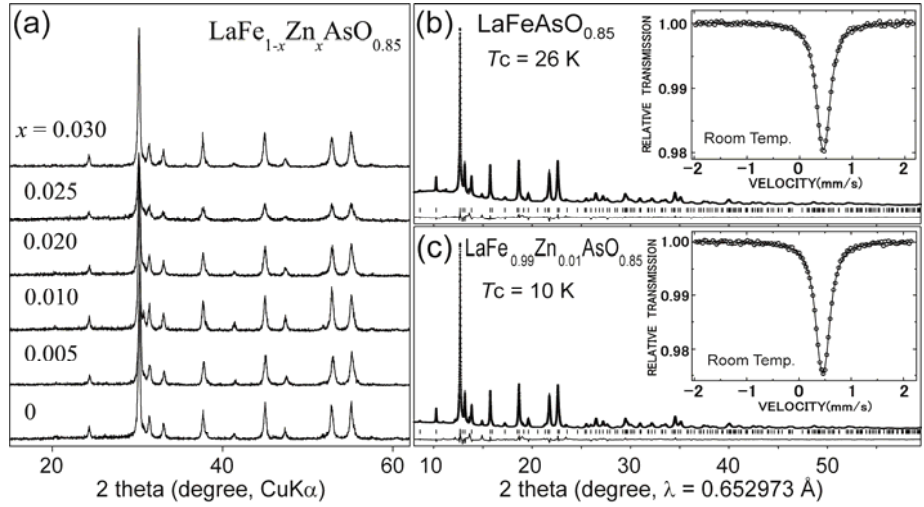

Fig. 1 (a) Powder XRD patterns of LaFe$_{1-x}$Zn$_x$AsO$_{0.85}$ ($x$ = 0 - 0.03). Peak identification for all is successful with *P4/nmm*. Rietveld analysis of the SXRD profiles for (b) LaFeAO$_{0.85}$ and (c) LaFe$_{0.99}$Zn$_{0.01}$AsO$_{0.85}$. The dots and the lines represent the observed and the calculated intensity, respectively. The difference between those is shown at the bottom each. The vertical bars indicated expected Bragg reflections. The inset shows the Mössbauer spectrum.

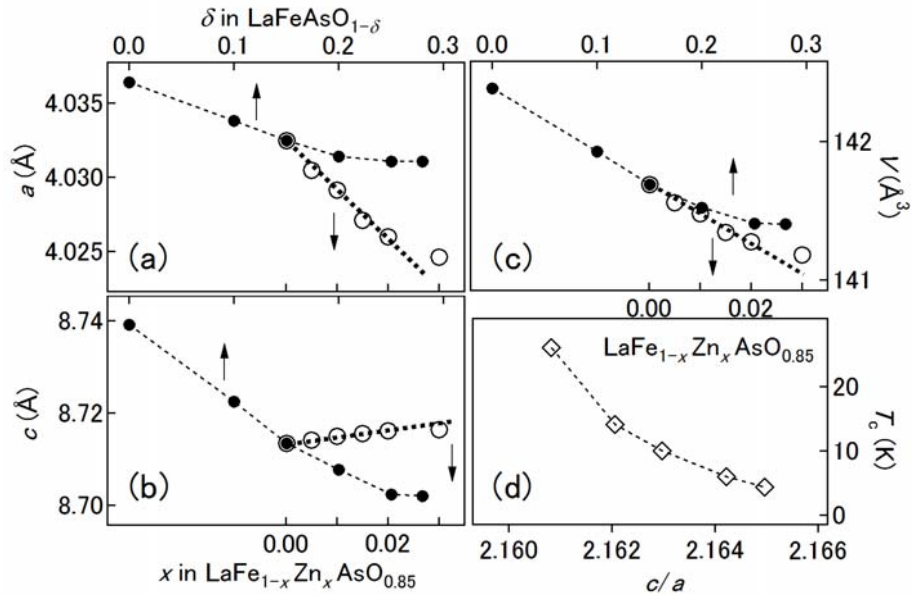

Fig. 2 (a-c) Evolution of the unit-cell dimensions of LaFe$_{1-x}$Zn$_x$AsO$_{0.85}$ (bottom axes) and LaFeAsO$_{1-\delta}$ (top axes, taken from ref. 20). (d) $T_c$ vs. $c/a$. The dashed lines are a guide to the eye.

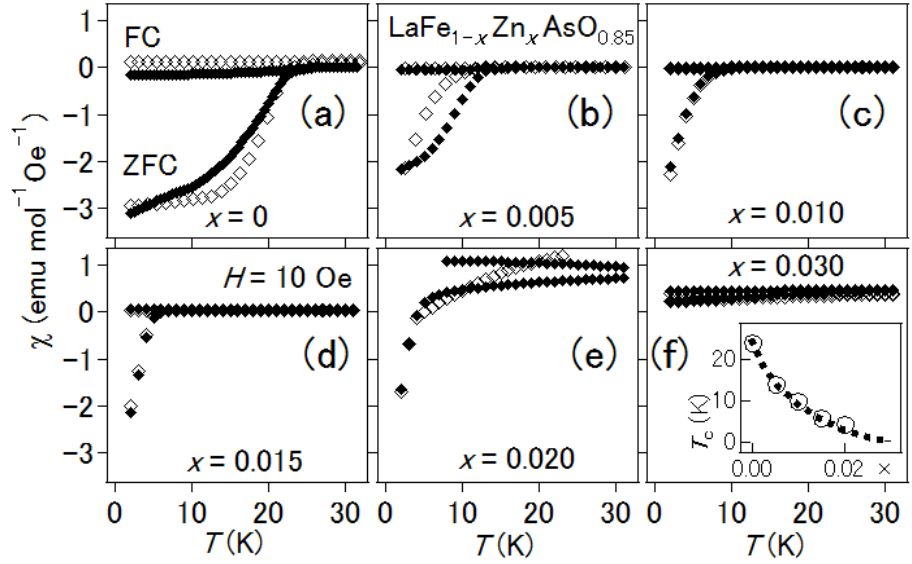

Fig. 3 (a-f) $T$ and $x$ dependence of $\chi$ of LaFe$_{1-x}$Zn$_x$AsO$_{0.85}$ measured at 10 Oe. The open and closed symboles respectively respsent two independent sets of data. The inset shows $T_c$ vs. $x$.

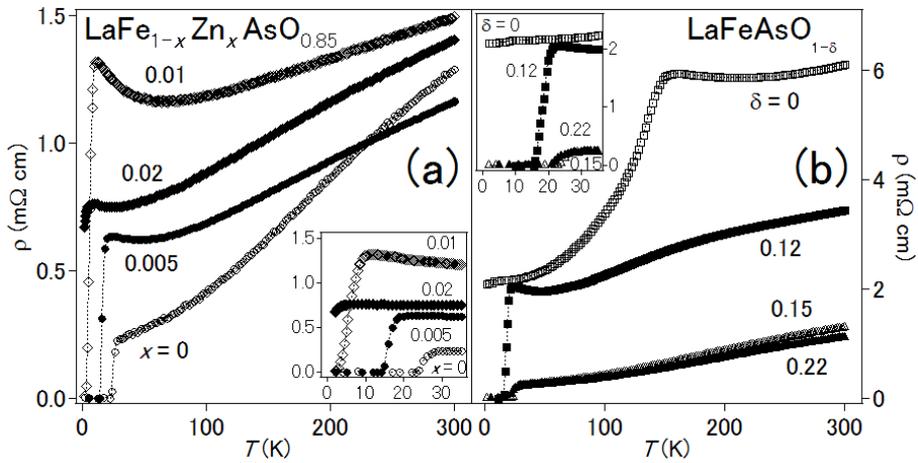

Fig. 4. $T$ and $x$ dependence of $\rho$ of (a) LaFe$_{1-x}$Zn$_x$AsO$_{0.85}$ and (b) LaFeAsO$_{1-\delta}$. Inset shows an expanded view.